\documentclass[english,fleqn,allpages]{ISTE-Standard}[2005/11/14]

\usepackage{natbib}

\usepackage{amsthm}

\newsavebox{\fminibox} \newlength{\fminilength} 

\usepackage[utf8]{inputenc}
\usepackage[english]{babel}
\usepackage{amssymb}
\usepackage{url}
\usepackage{dirtytalk} 

\usepackage{epstopdf}
\usepackage{subfigure}

\title[A review of the Dividend Discount Model]{A review of the Dividend Discount Model: from deterministic to stochastic models}

\makeatletter
\let\bbl@nonfrenchlistspacing\relax
\makeatother

\begin{document}

\raggedbottom



\mainmatter

\ChapterAuthor{A review of the Dividend Discount Model: from deterministic to stochastic models}{Guglielmo \Name{D'Amico} and Riccardo \Name{De Blasis}} \label{chap-struct}

\markboth{A review of the Dividend Discount Model}{A review of the Dividend Discount Model}

\section{Introduction}\label{Sec:Intro}
    This chapter presents a review of the dividend discount models starting from the basic models \citep{williams1938theory,gordon1956capital} to more recent and complex models \citep{ghezzi2003stock,barbu2017novel,dmultivariate} with a focus on the modelling of the dividend process rather than the discounting factor, that is assumed constant in most of the models. The Chapter starts with an introduction of the basic valuation model with some general aspects to consider when performing the computation. Then, Section \ref{Sec:Gordon} presents the Gordon growth model \citep{gordon1962investment} with some of its extensions \citep{malkiel1963equity,fuller1984simplified,molodovsky1965common,brooks1990n,barsky1993does}, and reports some empirical evidences. Extended reviews of the Gordon stock valuation model and its extensions can be found in \cite{kamstra2003pricing} and \cite{damodaran2012investment}. In Section \ref{Sec:Markov}, the focus is directed to more recent advancements which make us of the Markov chain to model the dividend process \citep{hurley1994realistic,yao1997trinomial,hurley1998generalized,ghezzi2003stock,barbu2017novel,dmultivariate}. The advantage of these models is the possibility to obtain a different valuation that depends on the state of the dividend series, allowing the model to be closer to reality. In addition, these models permit to obtain a measure of the risk of the single stock or a portfolio of stocks.
    
\section{General Model}\label{Sec:General}
    Stock valuation is one of the basic aspects of financial markets. Discussions about the fair price of a stock, or its overpricing and underpricing, have always been of paramount importance to investors. \cite{williams1938theory} was the first to recognise that market prices and fundamental values are \say{separate and distinct things not to be confused}. In his work, he states that an asset's intrinsic long-term value is the present value of all future cash flows, i.e., dividends and future selling price.
    
    Let $P(t)$ be the random variable giving the fundamental value of a stock at time $t\in \mathbb{N}$.  Let $D(t)$ be the dividend at time $t\in \mathbb{N}$, also assumed to be a random variable, and denote by $k_e(t)$ the required rate of return on the stock at time $t$. If we buy a stock at time $t$ and plan to sell it at time $t+1$, the price $p(t):=\mathbb{E}_{(t)}[P(t)]$ that we pay is the expected value of the stock price at time $t+1$ plus the cash flows distributed by the company, all discounted at an appropriate measure of risk $k_e(t)$,
    
    \begin{equation}\label{Eq:DDMonePeriod}
        p(t)=\mathbb{E}_{(t)}\Bigg[\frac{P(t+1)+D(t+1)}{1+k_e(t)}\Bigg],
    \end{equation}
    
    If we buy and hold the stock indefinitely, and assuming \citep[see, e.g.,][]{samuelson1973proof}
    
    \begin{equation}\label{Eq:BubbleCondition}
        \lim_{i\rightarrow +\infty}\mathbb{E}_{(t)}\Bigg[\frac{P(t+i)}{\prod_{j=0}^i\big[1+k_e(t+j)\big]}\Bigg]=0,
    \end{equation}
    then the price we pay is the expected value of all future cash flows in the form of dividends,
    
    \begin{equation}\label{Eq:DDM}
        p(t)=\sum_{i=0}^{+\infty}\mathbb{E}_{(t)}\Bigg[\frac{D(t+i+1)}{\prod_{j=0}^i\big[1+k_e(t+j)\big]}\Bigg].
    \end{equation}
    
    If condition (\ref{Eq:BubbleCondition}) is not assumed, then \cite{blanchard1982bubbles} proved that there could exist different solutions of the fundamental equation, i.e., there is the presence of bubbles in the stock market.
    
    To solve equation (\ref{Eq:DDM}), we have to identify two inputs, namely future dividends and the required measure of risk. When estimating future dividends, because of the impossibility of making predictions through to infinity, many models make assumptions about the dividend growth. The basic Gordon model \citep{gordon1962investment} is based on a constant dividend growth rate, while multistage models are advanced by \cite{brooks1990n} and \cite{barsky1993does} to better describe the dividend growth series. \cite{donaldson1996new} generalise the Gordon growth model to allow for arbitrary dividend growth and discount rates using a Monte Carlo simulation. On the contrary, others models apply specific stochastic processes to forecast dividends. \cite{gutierrez2004switching} propose a model which allows a regime switching in the dividend process and \cite{korn2005stocks} model dividends as a deterministic transformation of a Levy process. \cite{hurley2013calculating} introduces dividends modelled as a Bernoulli process with a continuous set of values, while \cite{eisdorfer2014pricing} model the time series behaviour of dividend growth rates with a first-order autoregressive process. In this Chapter we focus on how the various models make assumptions about the dividend process, with a particular attention to the Markov chain based models.
    
    The second input of the equation is the discount factor $k_e(t)$, or cost of equity, that represents a measure of the asset's riskiness. In most of the dividend discount models, it is assumed to be constant, $k_e$. Traditionally, the estimation of $k_e$ has been performed using the Capital Asset Pricing Model (CAPM). This model originates from the idea of mean-variance efficient portfolio of \cite{markowitz1952portfolio}, and it is formalised by \cite{sharpe1964capital} and \cite{lintner1965security} and extended by \cite{black1972capital}. The rationale of the model is that risky investments $R_{i}$, e.g., stocks in financial markets, are expected to be more remunerating than the risk-free assets
    \begin{align}
        & \mathbb{E}[R_i]=R_f+\beta_{im}(\mathbb{E}[R_m]-R_f),\\
        & \beta_{im}=\frac{Cov[R_i,R_m]}{Var[R_m]},
    \end{align}
    where $R_m$ is the return on the market portfolio, and $R_f$ is the return on the risk-free asset. The \cite{black1972capital} version substitutes the risk-free rate with a zero-beta portfolio uncorrelated with the market. The coefficient $\beta_{im}$ represents the correlation of the stock with the market, and can be estimated as the slope coefficient of the OLS regression
    \begin{equation}
        Z_{it} = \alpha_{im}+\beta_{im}Z_{mt}+\epsilon_{it},
    \end{equation}
    where $Z_{it}$ is the excess return of the stock on the risk-free asset, or equity premium, and $Z_{mt}$ is the market risk premium, $\mathbb{E}[R_i]-R_f$. In practical applications, the market return and the risk-free rate are proxied by market indices, e.g., S\&P 500 Index, and government treasury bonds, respectively. The estimation is based on a period of time that generally extends to about five years of historical data \citep{campbell1997econometrics}.
    
    Many authors provides empirical evidence on the CAPM application \citep[see, e.g.,][]{jensen1972capital,fama1973risk,blume1973new,basu1977investment,fama1992cross,fama1993common}, while \cite{roll1977critique} criticise it because the market portfolio is not observable and therefore the model is not testable. For a comprehensive description of the CAPM models and its variations with econometrics analysis see, e.g., \cite{campbell1997econometrics,cochrane2009asset}.
    
    In general, the dividend discount model is a very attractive model because it is intuitive and easy to implement. Nevertheless, it encounters much criticism because of the limits it poses. The main argument is the applicability of the model only to certain firms with stable, high-paying dividend policy. Moreover, the firms' recent practice of performing share buybacks instead of paying dividends, for obvious tax reasons, reduces the dividend cash flow and the application of the dividend discount model results in an underestimation of the value of the firm. The same principle applies to other assets that are ignored in the model, e.g., the value of brand names. However, share buybacks and values of other assets can be included in the dividends flow and treated as such with adequate adjustments \citep[see, e.g.,][]{damodaran2012investment}.
    
\section{Gordon Growth Model and Extensions}\label{Sec:Gordon}
    Equation (\ref{Eq:DDM}) can be rewritten in terms of dividend growth, defining
    
    \begin{equation}\label{Eq:DivGrowth}
        g(t)=\frac{D(t+1)-D(t)}{D(t)},
    \end{equation}
    as the growth rate of dividends from time $t$ to time $t+1$, so that $D(t+1)=D(t)(1+g(t))$ and $D(t+2)=D(t)(1+g(t))(1+g(t+1))$. Then, the price becomes,
    
    \begin{equation}\label{Eq:GeneralDivDiscount}
        p(t)=D(t)\sum_{i=0}^{+\infty}\mathbb{E}_{(t)}\Bigg[\prod_{j=0}^{i}\frac{1+g(t+j)}{1+k_e(t+j)}\Bigg].
    \end{equation}
    
    Assuming a constant dividend growth rate $g(t+j)=g$ and a constant discounting factor $k_e(t+j)=k_e$, equation (\ref{Eq:GeneralDivDiscount}) becomes
    
    \begin{equation}\label{Eq:GeneralDivDiscountConstant}
        p(t)=D(t)\sum_{i=0}^{+\infty}\frac{(1+g)^i}{(1+k_e)^i},
    \end{equation}
    and summing the geometric progression, we obtain the \textit{Gordon fundamental price estimate} \citep{gordon1962investment}
    
    \begin{equation}\label{Eq:GordonGrowthModel}
        p^G(t)=D(t)\frac{1+g}{k_e-g}, \quad\text{or}\quad p^G(t)=\frac{D(t+1)}{k_e-g}, 
    \end{equation}
    with the constraint $g<k_e$ to obtain a finite price.
    
    The Gordon model is straightforward because it requires only estimates of the dividend growth rate and discount rate, that are both easily obtained from a company's historical data. Nevertheless, it has some limitations. The model can result in incorrect estimations of the price when the growth rate approaches the discount rate, as the price tends to grow up to infinity. Therefore, this model is more suitable for companies with a stable dividend policy with a growth that is less than the growth of the economy. Moreover, empirical applications of the Gordon model show that dividends tend to grow  exponentially, meaning that a linear growth model is not suitable for the stock valuation \citep[see, e.g.,][]{campbell1987cointegration,west1988dividend}.

    The assumption of constant growth of the dividends forever is not realistic. To relax this assumption, \cite{malkiel1963equity} introduces a 2-stage model, with the first period of $n$ years of extraordinary growth followed by a stable growth forever. The value of a stock can be obtained as the sum of first years values, calculated from the general model plus a discounted value of the Gordon growth model at year $n$:
    \begin{equation}\label{Eq:2stageDDM}
        p^{2st}(t)=\mathbb{E}_{(t)}\Bigg[\sum_{i=0}^n\frac{D(t+i+1)}{\prod_{j=0}^i\big[1+k_e(t+j)\big]}+\frac{P^G(n)}{\prod_{j=0}^n\big[1+k_e(t+j)\big]}\Bigg],
    \end{equation}
    where $P^G(n)$ is the Gordon growth fundamental price estimate (\ref{Eq:GordonGrowthModel}) at year $n$.
    
    A further assumption of constant growth in the first phase, $g_h$, and constant discount rate $k_{e,h}$, simplifies equation (\ref{Eq:2stageDDM}) to
    \begin{equation}\label{Eq:2stageDDMconstant}
        p^{2st}(t)=\frac{D(t)(1+g_h)\bigg[1-\frac{(1+g_h)^n}{(1+k_{e,h})^n}\bigg]}{k_{e,h}-g}+\frac{P^G(n)}{(1+k_{e,h})^n},
    \end{equation}
    
    This model is suitable for valuing companies that expect to have an initial growth period higher than normal, because of a specific investment or a patent right, that will result in higher profits. At the same time, it presents some limits. First, the growth rate is expected to drop drastically from high to normal level, and second, it is hard to define the length of the high growth period in practical terms.
    
    To avoid the sharp drop from high to stable growth rate, \cite{fuller1984simplified} propose a linear decline of the growth in their \say{H} model. The high growth phase with decline is assumed to last $2H$ periods up to the stable growth phase $g_n$, with an initial growth rate $g_a$. The model assumes that the discount rate $k_e$ is constant over time, as well as the dividend payout ratio.
    
    \begin{equation}\label{Eq:H-DDM}
        p^{H}(t)=\frac{D(t)(1+g_a)}{k_e-g_n}+\frac{D(t)H(g_a-g_n)}{k_e-g_n},
    \end{equation}
    
\begin{figure}[t]
        \centering
        \includegraphics[width=0.9\columnwidth]{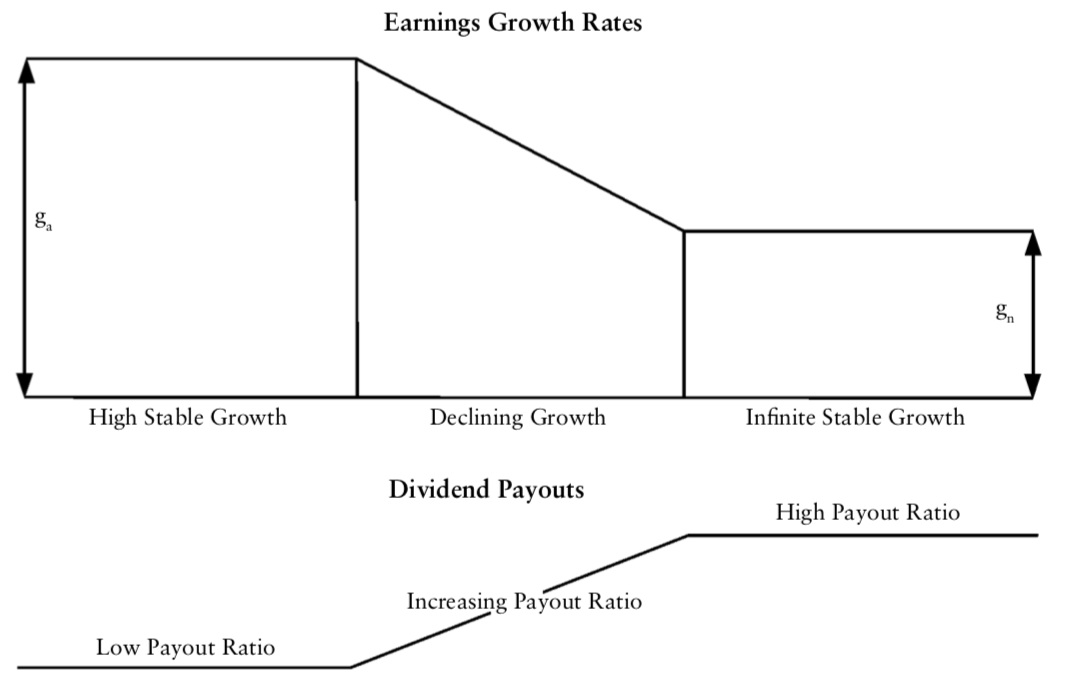}
        \caption[Expected growth in a 3-stage dividend discount model]{Expected growth in a 3-stage dividend discount model \citep[Figure from][p. 341]{damodaran2012investment}}
        \label{fig:3-stage}
    \end{figure}

    A constant payout ratio assumption poses some limits to this model. Generally, a company is expected to have lower payout ratios in high growth phases and higher payout ratios in the stable growth phase, as shown in Figure \ref{fig:3-stage}.
    
    A 3-stage model, initially formulated by \cite{molodovsky1965common} and derived from a combination of the H model and the 2-stage model, with the inclusion of a variable payout policy and different discount factors for the various phases, overcomes the limits of previous models, but it requires a larger number of inputs. Let $k_{e,h}$, $k_{e,d}$, and $k_{e,st}$ be the discount factors for high, declining, and stable phases, respectively. Let $g_a$ and $g_n$ be the growth rate at the beginning and the end of the period. Let $EPS$ be the earnings per share, and $\Pi_a$ and $\Pi_n$ the payout ratios at the beginning and end of the period, respectively. The stock valuation for the 3-stage model is
    
    \begin{equation}\label{Eq:3stageDDM}
        \begin{aligned}
        p^{3st}(t)=\sum_{i=0}^{n1}\frac{EPS(t)\Pi_a(1+g_a)^i}{(1+k_{e,h})^i}+\sum_{i=n1+1}^{n2}\frac{D(t+i)}{(1+k_{e,d})^i}\\
        +\frac{EPS(t+n2)\Pi_n(1+g_n)}{(k_{e,st}-g_n)(1+k_{e,h})^{n1}(1+k_{e,d})^{n2-n1}},
        \end{aligned}
    \end{equation}
    
    \vspace{1em}
    An empirical comparison of the Gordon model and its variations is in \cite{sorensen1985some}. The authors analyse the intrinsic value of a random sample of 150 firms from the S\&P 400 using data available in 1981, from four different valuation models, price/earning model, constant growth model, two-period, and three-period model. They base the analysis on normalised earnings and a dividend payout ratio of approximately 45 per cent. The discount factor is calculated using the CAPM model for the growth period, according to the beta of the stock and the high growth period is assumed to last five years for all the stock. Then, based on the assumption that all mature firms look alike, an equal risk measure of 8\% among all the stocks is adopted for the stable phase.
    
    For every model, the authors generate five portfolios of 30 stocks each, ordered from undervalued to overvalued securities, estimating returns for two years. Results show that the increased complexity of the model improves the annualised returns. As well as looking at the risk characteristics of the portfolios, the 3-stage model outperforms the other model.
    
    \cite{brooks1990n} generalise the 2-stage model from \cite{malkiel1963equity}. They propose an N-stage model, with quarterly dividends and fractional periods. Within each stage, dividends growth is assumed constant, and the discount rate is based on quarterly compounding $r_e = (1-k_e)^{\frac{1}{4}}-1$. They test the model on the case of Commonwealth Edison Company (CWE), an electricity supplier, estimating the required rate of return for three cases: (a) annual dividends, no fractional period; (b) quarterly dividends, no fractional periods; and (c) quarterly dividends, fractional periods. They show that ignoring quarterly compounding and fractional periods the results present a downward bias.

    Another extension of the Gordon growth models is given in \cite{barsky1993does}. The authors propose to model the permanent dividend growth as a geometric average of past dividend changes:
    \begin{equation}\label{Eq:IMA}
        g(t)=(1-\theta)\sum_{i=0}^t\theta^i \Delta D(t-i) + \theta^tg(0)
    \end{equation}
    with $g(t)$ following a random walk process and, thus, change in dividends following an IMA(1,1).
    
    \cite{donaldson1996new} generalise the Gordon growth model allowing for arbitrary dividend growth and discount rates. Their methodology involves a Monte Carlo simulation and numerical integration of the random joint process of dividend growth and discount rates
    
    \begin{equation}\label{Eq:DKprocess}
        y(t+1)=\frac{1+g(t+j)}{1+k_e(t+j)}.
    \end{equation}
    
    They forecast a range of possible evolution of the process $y(t+1)$ up to a certain point in the future, $t+I$, and calculate the average of several estimations of the present stock value
    
    \begin{equation}\label{Eq:DKprice}
        p(t) =  D(t)\sum_{i=0}^I\prod_{j=0}^iy(t+1).
    \end{equation}
    
\section{Markov chain stock models}\label{Sec:Markov}
    According to equation (\ref{Eq:GeneralDivDiscount}), the stock valuation is obtained through two inputs, namely the dividend growth and the discount factor. The idea of the Markov chain stock models is to describe the dividend growth rate as a sequence of independent, identically distributed, discrete random variables, and model it as a Markov process.  In all these models, the discount factor $k_e$ is kept constant.
    
    \cite{hurley1994realistic} model the dividend growth as a \textit{Markov dividend stream}. They assume that in each period the dividend can increase with probability $q$, be the same with probability $1-q$, to resemble a step pattern in the long term. Moreover, they include the possibility for the firm to go bankrupt, with probability $q_B$. They propose two variations of the model, an additive model and a geometric model, both giving an estimation of the value, along with a lower bound estimation for each of these values. 
    
    In the \textit{additive model}, the dividend at time $t+1$ increase by the amount $\Delta$ with probability $q$, and assuming a constant discount rate $k_e$, the value of the firm is
    \begin{equation}\label{Eq:HurleyAdditiveValue}
        p(t)=\begin{cases}
        D(t) + \Delta + p(t+1)\frac{D(t)+\Delta}{1+k_e} &\mbox{with prob } q\\
        D(t) + p(t+1)\frac{D(t)}{1+k_e} &\mbox{with prob } 1-q-q_B\\
        0 &\mbox{with prob } q_B\\
        \end{cases},
    \end{equation}
    and the closed form solutions for the value and the lower bound are
    
    \begin{equation}\label{Eq:HurleyAddClosed}
        p^A(t)=\frac{D(t)}{k_e}+\bigg[\frac{1}{k_e}+\frac{1}{k_e^2}\bigg] q\Delta,
    \end{equation}
    and
    
    \begin{equation}\label{Eq:HurleyAddLower}
        p^{A}_{low}(t)=\frac{D(t)(1-q_B)}{k+q_B}+\bigg[\frac{1}{k+q_B}+\frac{1}{(k+q_B)^2}\bigg] q\Delta.
    \end{equation}
    Note that, when $q_b=0$, $p^{A}_{low}=p^A$.
    
    The \textit{geometric model} assumes that the dividend increases with a growth rate $g$ and with a probability $q$
    \begin{equation}\label{Eq:HurleyGeometricDividend}
        D(t+1)=\begin{cases}
        D(t)(1+g) &\mbox{with prob } q\\
        D(t)  &\mbox{with prob } 1-q-q_B
        \end{cases}.
    \end{equation}
    The closed form solutions for the value and the lower bound become
    
    \begin{equation}\label{Eq:HurleyGeoClosed}
        p^G(t)=\frac{D(t)(1+qg)}{k_e-qg},
    \end{equation}
    and
    
    \begin{equation}\label{Eq:HurleyGeoLower}
        p^{G}_{low}(t)=D(t)\bigg[\frac{1+qg-q_B}{k_e-(qg-q_B)}\bigg].
    \end{equation}
    
    It is worth noting that the geometric model reduces to the Gordon model, setting the expected growth rate to $qg-q_B$, or, if we exclude the possibility of bankruptcy, setting the expected growth rate to $qg$.
    
    An empirical application to three stocks, provided in \cite{hurley1994realistic}, shows that the geometric method performs well when the dividend series is erratic and does not always show increases. The model gives an estimation that is very close to the actual stock prices.
    
    \cite{hurley1998generalized} formulate a generalised version of their model to include the possibility of a decrease in the dividends, so the dividend at time $t$ is $D(t)=D(t-1)+\Delta_i$ for the additive model, and $D(t)=D(t-1)(1+g_i)$ with probability $q_i$ for the geometric model. Both $\Delta_i$ and $g_i$ include the possibility of dividends reduction, or suspensions. Under the condition $q_0+\sum_{i=1}^nq_i=1$, the closed form solution for both models are
    
    \begin{equation}\label{Eq:HurleyAddGeneral}
        p^A(t)=\frac{D(t)}{k_e}+\bigg[\frac{1}{k_e}+\frac{1}{k_e^2}\bigg] \sum_{i=1}^nq_i\Delta_i,
    \end{equation}
    and
    
    \begin{equation}\label{Eq:HurleyGeoGeneral}
        p^G(t)=D(t)\frac{1+\sum_{i=1}^nq_ig_i}{k_e-\sum_{i=1}^nq_ig_i}.
    \end{equation}
    
    When $n=1$, the models reduces to \cite{hurley1994realistic} models.
    
    \cite{yao1997trinomial} advances the same proposal of a dividend reduction extending \cite{hurley1994realistic} models. The author introduces a trinomial dividend valuation model and extends the additive model, where the dividend at time $t+1$ is 
    
    \begin{equation}\label{Eq:YaoAdditiveDividend}
        D(t+1)=\begin{cases}
        D(t) + \Delta  &\mbox{with prob } q^u\\
        D(t) - \Delta  &\mbox{with prob } q^d\\
        D(t)  &\mbox{with prob } q^c=1-q^u-q^d
        \end{cases},
    \end{equation}
    with closed solution for the stock value
    \begin{equation}\label{Eq:YaoAddValue}
        p^A(t)=\frac{D(t)}{k_e}+\bigg[\frac{1}{k_e}+\frac{1}{k_e^2}\bigg] (q^u-q^d)\Delta.
    \end{equation}
    Then, the geometric model, with 
    
    \begin{equation}\label{Eq:YaoGeometricDividend}
        D(t+1)=\begin{cases}
        D(t)(1+g)  &\mbox{with prob } q^u\\
        D(t)(1-g)  &\mbox{with prob } q^d\\
        D(t)  &\mbox{with prob } q^c=1-q^u-q^d
        \end{cases},
    \end{equation}
    and closed solution
    
    \begin{equation}\label{Eq:YaoGeoValue}
        p^G(t)=D(t)\frac{1+(q^u-q^d)g}{k_e-(q^u-q^d)g}.
    \end{equation}
    
    Lower bounds for both models are also given by the author. Moreover, a practical application on five firms, provided in \cite{yao1997trinomial}, shows that the model produces better estimates than \cite{hurley1994realistic}.
    
    \cite{ghezzi2003stock} start from the previous Markov models to formulate a more general Markov chain stock model. The authors begin with a description of the simple model for the dividend growth rate using a 2-state discrete Markov chain, and a constant discount rate $r=1+k_e$. Finally, they extend the model to an n-state Markov chain and define a vector of price-dividend ratios as the solution of a system of linear equations.
    
    In previous models, \cite{hurley1994realistic,hurley1998generalized} and \cite{yao1997trinomial} assume that the dividend growth rates are independent, identically distributed, discrete random variables, thus obtaining one closed form solution irrespective of the state of the dividend. On the contrary, \cite{ghezzi2003stock} relax the  i.i.d. assumption and obtain a different price-dividend solution for each state of the dividends. This variety allows the Markov chain stock model to be closer to reality.
    
    The dividend series obeys the difference equation
    
    \begin{equation}\label{Eq:GhezziDividend}
        D(k+1)=G(k+1)D(k), \quad k=t,t+1,\hdots,
    \end{equation}
    where $G(k+1)$ is the dividend growth factor described by a Markov chain. 
    
    The dividend series relation (\ref{Eq:GhezziDividend}) states that given the initial dividend value $D(0)=d\in \mathbb{R}$, we can compute the next random dividend $D(1)=G(1)D(0)=G(1)d$, and the next $D(2)=G(2)D(1)=G(2)G(1)d$, and so on. Generally $D(n)=\prod_{i=1}^{n}G(i)d$.
    
    The combination of the dividend discount model equation (\ref{Eq:GeneralDivDiscount}) and (\ref{Eq:GhezziDividend}), with a constant discount factor $r$, i.e., one plus the required rate of return, yields
    \begin{equation}\label{Eq:GhezziValue}
        p(k)=d(k)\sum_{i=1}^{+\infty}\frac{\mathbb E_{(k)}[\prod_{j=1}^{i}G(k+j)]}{r^{i}} =: d(k)\psi_1(g(k)),
    \end{equation}   
    where $d(k)$ and $g(k)$ are the values at time $k$ of the dividend process and of the growth dividend process, respectively, and $\psi_1(g(k))$ is the price-dividend ratio.
    
    The simple case is modelled with a 2-state Markov chain taking values in the state space $E=\{g_1,g_2\}$. Let $\mathbf{P}=(p_{ij})_{i,j\in E}$ be the one-step transition probability matrix of this Markov chain, and let
    
    \begin{equation} \label{Eq:GhezziCondition}
        {\bf A1}: \overline{g}: = \max(p_{11}g_{1}+p_{12}g_{2},p_{21}g_{1}+p_{22}g_{2})<r,
    \end{equation}
    be the largest one step conditional expectation on the dividend growth rate.
    
    If A1 holds true, then the series $p(k)=\sum_{i=1}^{+\infty}\frac{\mathbb E_{(k)}[D(k+i)]}{r^{i}}$
    converges and satisfies the asymptotic condition in (\ref{Eq:BubbleCondition}), and the pair $(\psi_1(g_{1}),\psi_1(g_{2}))$ is the unique and non-negative solution of the linear system 
    \begin{equation}\label{Eq:Ghezzi2stateSystem}
    \begin{aligned}
    & \psi_1(g_{1})=p_{11}\frac{\psi_1(g_{1})g_{1}+g_{1}}{r}+p_{12}\frac{\psi_1(g_{2})g_{2}+g_{2}}{r}\\
    & \psi_1(g_{2})=p_{21}\frac{\psi_1(g_{1})g_{1}+g_{1}}{r}+p_{22}\frac{\psi_1(g_{2})g_{2}+g_{2}}{r}.
    \end{aligned}
    \end{equation}
    
    Assuming that for any given $D(k)$ we obtain the same $\mathbb{E}[D(t+1)]$, irrespective of the initial states $g_1,g_2$, then $p_{11}=q$ and $p_{22}=1-q$, therefore the solution to (\ref{Eq:Ghezzi2stateSystem}) becomes
    
    \begin{equation}\label{Eq:GhezziReduction}
        \psi_1(g_{1})=\psi_1(g_{2})=\frac{qg_1+(1-q)g_2}{r-1g_1-(1-q)g_2},
    \end{equation}
    thus implying that the same price-dividend ratio is attached to each state, sharing the same results as \cite{hurley1994realistic,hurley1998generalized} and \cite{yao1997trinomial}.
    
    Results can be easily extended to the case of an s-state Markov chain with state space $E=\{g_{1},g_{2}, \ldots, g_{s}\}$, where assumption A1 becomes,
    \begin{equation}\label{Eq:GhezziConditionNstate}
        \overline{g}:=\max_{i\in E}\bigg(\sum_{j=1}^{s}p_{ij}g_{j}\bigg)<r.
    \end{equation}
    
    If $\overline{g}<r$ the series (\ref{Eq:GhezziValue}) converges and the unique and non-negative solution to the linear system is
    
    \begin{equation}\label{Eq:GhezziNstateSystem}
    \psi(g_{i})=\sum_{j=1}^s p_{ij}\frac{\psi(g_{j})g_{j}+g_{j}}{r}, \quad i=1,2,\hdots,s.
    \end{equation}
    
    This model has the advantage of assigning a different price-dividend ratio to each value of the states, that does not depend on the time. Forecasts on the dividend growth rate are updated based on the previous value of the state, according to the Markov property, thus the price of the stock is updated according to the state of the dividend process. On the contrary, all previous models make fixed assumptions on forecasts and obtain a unique valuation.
    
    \cite{agosto2015variance} complement the model calculating a closed-form expression for the variance of random stock prices in a multinomial setting. The authors argue that for proper investment decisions a measure of risk should be taken into consideration. Thus applying the standard mean-variance analysis, an investor can deal with financial decisions under uncertainty. In their model, they relate the variance of stock prices with the variance of the dividend rate of growth, obtaining a measure of the stock riskiness.
    
    An extension of the Markov stock model is given in \cite{barbu2017novel}. The authors provide a formula for the computation of the second order moment of the fundamental price process in the case of a 2-state Markov chain,
    \begin{equation}\label{Eq:BarbuSeries}
        \begin{aligned}
        & p^{(2)}(k)=\sum_{i=1}^{+\infty}\frac{\mathbb{E}_{k}[D^{2}(k+i)]}{r^{2i}}+ \\
        & \qquad 2\sum_{i=1}^{+\infty}\sum_{j>i}\frac{\mathbb{E}_{k}[D(k+i)D(k+j)]}{r^{i+j}}=\psi_{2}(g(k)) \: d^{2}(k).
        \end{aligned}
    \end{equation}
    
    To obtain the convergence of the series (\ref{Eq:BarbuSeries}) and to satisfy the asymptotic condition in (\ref{Eq:BubbleCondition}) and 
    \begin{equation*}
        lim_{N\rightarrow +\infty} \sum_{i=1}^{N}\frac{\mathbb{E}_{k}[D(k+i)P(k+N)]}{r^{i+N}}=0,
    \end{equation*}
    the authors introduce a further assumption that avoids the presence of speculative bubbles,
    \begin{equation}\label{Eq:BarbuCondition}
        {\bf A2}: \overline{g}^{(2)}:=max(p_{11}g_{1}^{2}+p_{12}g_{2}^{2},p_{21}g_{1}^{2}+p_{22}g_{2}^{2})<r^{2},
    \end{equation}
    where $\overline{g}^{(2)}$ is the largest one step second order moment of the dividend growth rate.
    
    If assumptions A1 and A2 hold true, the pair $(\psi_{2}(g_{1}),\psi_{2}(g_{2}))$ is the unique and nonnegative solution of the linear system 
    \begin{equation}\label{Eq:Barbu2stateSystem}
        \begin{aligned}
            & \psi_{2}(g_{1})\big(r^{2}-p_{11}g_{1}^{2}\big)-\psi_{2}(g_{2})p_{12}g_{2}^{2}=p_{11}g_{1}^{2}\big(1+2\psi_{1}(g_{1})\big)+p_{12}g_{2}^{2}\big(1+2\psi_{1}(g_{2})\big)\\
            & \psi_{2}(g_{2})\big(r^{2}-p_{22}g_{2}^{2}\big)-\psi_{2}(g_{1})p_{21}g_{1}^{2}=p_{21}g_{1}^{2}\big(1+2\psi_{1}(g_{1})\big)+p_{22}g_{2}^{2}\big(1+2\psi_{1}(g_{2})\big).
        \end{aligned}
    \end{equation}

    To extend the results to an s-state Markov chain with state space $E=\{g_{1},g_{2}, \ldots, g_{s}\}$,  assumptions A1 should be formulated as (\ref{Eq:GhezziConditionNstate}) and A2 as follows:
    \begin{equation}\label{Eq:BarbuConditionNstate}
        \overline{g}^{(2)}:=\max_{i\in E}\bigg(\sum_{j=1}^{s}p_{ij}g_{j}^{2}\bigg)<r^{2}.
    \end{equation}

    In this general case, the systems (\ref{Eq:Ghezzi2stateSystem}) and  (\ref{Eq:Barbu2stateSystem}) can be conveniently represented in matrix form,
    \begin{equation}\label{matrixsystem_Psi}
        \left(\mathbf{I}_{r}-\mathbf{P} \cdot \mathbf{I}_{\mathbf{g}}\right)\cdot \mathbf{\Psi}_{1}=\mathbf{P}\cdot \mathbf{g},
    \end{equation} 
    \begin{equation}\label{matrixsystem}
        \bigg(\mathbf{I}_{r}^{2}-\mathbf{P} \cdot \mathbf{I}_{\mathbf{g}}^{2}\bigg)\cdot \mathbf{\Psi}_{2}=\mathbf{P}\cdot \left(\big(\mathbf{g} \diamond \mathbf{g}\big)+2\mathbf{\Psi}_{1}\diamond \big(\mathbf{g} \diamond \mathbf{g}\big)\right),
    \end{equation}
    where:
    \begin{itemize}
        \item $\mathbf{\Psi}_{1}=(\psi_{1}(g_{1}), \ldots,\psi_{1}(g_{n}))^{\top}$ and $\mathbf{\Psi}_{2}=(\psi_{2}(g_{1}), \ldots, \psi_{2}(g_{n}))^{\top}$,
        \item $\mathbf{I}_{r}:=r \mathbf{I}$, for any $ r \in \mathbb R^{*}:=\mathbb{R}- \{0\}$, and, more generally, $\mathbf{I}_{r}^n = \mathbf{I}_{r^n}$,
        \item $
                \mathbf{I}_{\mathbf{g}}=\label{eq_Ig}
                (I_{g}(i,j))_{i,j\in E},\,\, I_{\mathbf{g}}(i,j)= \left\{ {\begin{array}{ll}
                   g_{i},\,\, \text{if}\,\,i=j  \\
                   0,\,\, \text{if}\,\,\,\,i\neq j
                 \end{array} } \right. 
                $,
        \item $\mathbf{I}$ is the identity matrix of dimension $s \times s$,
        \item $\cdot$ denotes the row by column matrix product and $\diamond $ denotes the Hadamard element by element product.
    \end{itemize}
    
    The matrix $\left(\mathbf{I}_{r}-\mathbf{P} \cdot \mathbf{I}_{g}\right)$ is invertible, therefore the system (\ref{matrixsystem_Psi}) has a unique solution,
    \begin{equation}\label{matrixsystem_Psi_sol}
        \mathbf{\Psi}_{1}=\left(\mathbf{I}_{r}-\mathbf{P} \cdot \mathbf{I}_{\mathbf{g}}\right)^{-1} \cdot \mathbf{P} \cdot \mathbf{g}.
    \end{equation}

    Similarly, the matrix 
    $\left(\mathbf{I}_{r}^{2}-\mathbf{P} \cdot \mathbf{I}_{g}^{2}\right)$ is invertible and the solution to the system (\ref{matrixsystem}) is
    \begin{eqnarray}
        \mathbf{\Psi}_{2}&=& \nonumber \left(\mathbf{I}_{r}^{2}-\mathbf{P} \cdot \mathbf{I}_{\mathbf{g}}^{2}\right)^{-1} \cdot \mathbf{P}\cdot \left(\big(\mathbf{g} \diamond \mathbf{g}\big)+2\mathbf{\Psi}_{1}\diamond (\mathbf{g} \diamond \mathbf{g})\right)\\
        &=& \label{matrixsystem_Psi2_sol}\left(\mathbf{I}_{r^{2}}-\mathbf{P} \cdot \mathbf{I}_{\mathbf{g}^{2}}\right)^{-1} \cdot \mathbf{P}\cdot \left(\big(\mathbf{g} \diamond \mathbf{g}\big)+2\mathbf{\Psi}_{1}\diamond \big(\mathbf{g} \diamond \mathbf{g}\big)\right),
    \end{eqnarray}
    
    Relation (\ref{matrixsystem_Psi2_sol}) represents an explicit formula for the second-order price-dividend ratio, that multiplied by $d^{2}(t)$ results in the second moment of the price process that is expressed in function of the model parameters $\mathbf{P}$ and $\mathbf{g}$.
    
    \cite{barbu2017novel} completed the Markov stock model framework developing non parametric statistical techniques for the inferential analysis of the model where they  propose estimators of price, risk and forecasted prices. For each estimator they demonstrate that they are strongly consistent and that, after proper centralisation and normalisation, they converge in distribution to normal random variables. Finally, they give the interval estimators.
    
    A further generalisation of \cite{ghezzi2003stock} is available in \cite{d2013semi}. The author models the dividend growth rate as a semi-Markov chain. In this setting, prices become duration dependent. Therefore, they are influenced by the current state of the dividend growth process and by the elapsed time in the state. The same author proposes another extension of the model describing the dividend growth series via a continuous state space semi-Markov model \citep{d2017stochastic}.
    
\subsection{Multivariate Markov chain stock model}\label{Sec:Multivariate}
    
    The previous analysis of the dividend discount model focused on the valuation of a single firm based on its dividend process. In this section, we analyse the problem of valuating multiple stocks when they constitute a financial portfolio. When dealing with more than one price series, it is important to consider the possible dependencies that characterise the pool of stocks. In a recent paper, \cite{agosto2018stochastic} compute the covariance between two stocks that may be held in a portfolio. They consider a Markov chain with state space equal to the set of possible couples of the growth-dividend values for both stocks. However, this strategy cannot be easily implemented in real applications, especially when we introduce dependencies between more than two stocks as the number of parameters to estimate increase drastically. 
    
    \cite{dmultivariate} propose and extension of the Markov stock model to a multivariate setting, computing the first and the second order price-dividend ratios. Moreover, the authors provide a formula for the computation of the variances and covariances between stocks in a portfolio. The model belongs to the class of mixture transition distribution models originated by \cite{raftery1985model} in a high order Markov chain setting and further extended in \cite{ching2006markov} to a multivariate Markov chain setting. This approach permits to overcome the limitations of \cite{agosto2018stochastic} because it reduces the number of parameters to estimate.
    
    With a portfolio of multiple stocks, $\alpha=1,2,\hdots,\gamma$, the dividend series expressed in (\ref{Eq:GhezziDividend}) becomes
    \begin{equation}\label{Eq:MultivariateDividendSeries}
        D^{(\alpha)}(k+1)=G^{(\alpha)}(k+1)\cdot D^{(\alpha)}(k),
    \end{equation}
    where $\{G^{(\alpha)}\}_{k\in \mathbb{N}}$ is the growth-dividend random process for stock $\alpha$, and the multivariate Markov chain model follows the relationship,
    \begin{equation}\label{Eq:MTDmodel}
        \mathbf{A}^{(\alpha)}(k+1)=\sum_{\beta=1}^{\gamma}\mathbf{A}^{(\beta)}(k)\cdot \lambda_{\beta,\alpha}\cdot \mathbf{P}^{(\beta,\alpha)},
    \end{equation}
    where:
    \begin{itemize}
        \item $\mathbf{A}^{\alpha}(k):=[A_{1}^{(\alpha)},\hdots,A_{m}^{(\alpha)}]$ is a probability distribution vector with $A_{i}^{(\alpha)}(k):=\mathbb{P}[G^{(\alpha)}(k)=i]$ being the probability of growth-dividend of stock $\alpha$ to be at time $k$ in state $i$,
        \item $\lambda_{\beta,\alpha}\in [0,1]$, $\sum_{\beta=1}^{\gamma}=1$,
        \item $\mathbf{P}^{(\beta,\alpha)}$ is the transition probability matrix of stock $\alpha$ given the state occupied one time step before by stock $\beta$, i.e.
    \end{itemize}
    \begin{equation}\label{Eq:MultiTransitionProb}
        \mathbf{P}^{(\beta,\alpha)}_{i,j}=\mathbb{P}[G^{(\alpha)}(k+1)=j \mid G^{(\beta)}(k)=i].
    \end{equation}
   
    According to equation (\ref{Eq:MTDmodel}) the probability distribution function of the growth-dividend process at time $k+1$ for the stock $\alpha$ depends on the state of the growth-dividend process of the same stock at time $k$, and, at the same time, on the set of states visited by each stock in the portfolio at time $k$.
    
    To extend the model to a multivariate setting, the price process series (\ref{Eq:GhezziValue}) and (\ref{Eq:BarbuSeries}) can be rewritten as,
    \begin{equation}\label{Eq:MultivariateSeries1}
        \begin{aligned}
            p^{(\alpha)}(\bm{g}(k))&=\sum_{i=1}^{+\infty}\frac{\mathbb{E}_{(k)}[D^{(\alpha)}(k+i)]}{r_{\alpha}^{i}}\\
            &=\sum_{i=1}^{+\infty}\Big(\frac{\mathbb{E}_{(k)}[\prod_{j=1}^{i}G^{(\alpha)}(k+j)]}{r_{\alpha}^{i}}\Big)d^{(\alpha)}(k),
        \end{aligned}
    \end{equation}
    \begin{equation}\label{Eq:MultivariateSeries2}
        \begin{aligned}
            p_{2}^{(\alpha,\beta)}(k)&:=\sum_{i=1}^{+\infty}\mathbb{E}_{(k)}\Big[\frac{D^{(\alpha)}(k+i)D^{(\beta)}(k+i)}{r_{\alpha}^{i}\cdot r_{\beta}^{i}}\Big]\\
            & +\sum_{i=1}^{+\infty}\sum_{j>i}\mathbb{E}_{(k)}\Big[\frac{D^{(\alpha)}(k+i)D^{(\beta)}(k+j)}{r_{\alpha}^{i}\cdot r_{\beta}^{j}}\Big]\\
            & + \sum_{i=1}^{+\infty}\sum_{j>i}\mathbb{E}_{(k)}\Big[\frac{D^{(\alpha)}(k+j)D^{(\beta)}(k+i)}{r_{\alpha}^{j}\cdot r_{\beta}^{j}}\Big],
        \end{aligned}
    \end{equation}
    respectively. Equation (\ref{Eq:MultivariateSeries2}) represents the fundamental formula of the price-product and reduces to the second order moment of the price process when considering the same price series, $\alpha=\beta$.
    
    To guarantee the convergence of the series (\ref{Eq:MultivariateSeries1}) and (\ref{Eq:MultivariateSeries2}) in the multivariate setting, \cite{dmultivariate} extend the transversality conditions in (\ref{Eq:GhezziConditionNstate}) and (\ref{Eq:BarbuConditionNstate}),
    \begin{align}
        & \overline{g}^{(\alpha;1)}:=\max_{\bm{e}^{(1)},\hdots,\bm{e}^{(\gamma)}}\Bigg(\sum_{j=1}^{m}\sum_{\beta=1}^{\gamma}\sum_{h=1}^{m}e_{h}^{(\beta)}\lambda_{\beta,\alpha}\mathbf{P}_{h,j}^{(\beta,\alpha)}g_{j}\Bigg)<r_{\alpha}, \label{Eq:MultivariateConditions1}\\
        & \overline{g}^{(\alpha;2)}:=\max_{\bm{e}^{(1)},\hdots,\bm{e}^{(\gamma)}}\Bigg(\sum_{j=1}^{m}\sum_{\beta=1}^{\gamma}\sum_{h=1}^{m}e_{h}^{(\beta)}\lambda_{\beta,\alpha}\mathbf{P}_{h,j}^{(\beta,\alpha)}(g_{j})^{2}\Bigg)<r_{\alpha}^{2}. \label{Eq:MultivariateConditions2}
    \end{align}
    
    If assumptions in (\ref{Eq:MultivariateConditions1}) holds true, then the first order price-dividend ratio, $\psi_{1}^{(\alpha)}(\bm{g}(k))$, can be computed as a linear system of $m^{\gamma}$ equations in $m^{\gamma}$ unknown that admits a unique solution,
    \begin{equation}\label{Eq:MultivariateFirstOrder}
    \begin{aligned}
        \psi_{1}^{(\alpha)}(g_{a_{1}}^{(1)},\ldots,g_{a_{\gamma}}^{(\gamma)})=\frac{1}{r_{\alpha}}\Big\{\sum_{j_{\alpha}=1}^{m}\sum_{\beta=1}^{\gamma}\sum_{h=1}^{m}e_{h}^{(\beta)}(k)\lambda_{\beta,\alpha}\mathbf{P}_{h,j_{\alpha}}^{(\beta,\alpha)}g_{j_{\alpha}}^{(\alpha)}+\\
        \sum_{j_{1},\hdots,j_{\gamma}=1}^{m}\psi_{1}^{(\alpha)}(g_{j_{1}}^{(1)},\hdots,g_{j_{\gamma}}^{(\gamma)})\cdot g_{j_{\alpha}}^{(\alpha)}\cdot \prod_{f=1}^{\gamma}\sum_{w=1}^{\gamma}\sum_{c=1}^{m}e_{c}^{(w)}(k)\lambda_{w,f}\mathbf{P}_{c,j_{f}}^{(w,f)}\Big\}.
    \end{aligned}
    \end{equation}
    
    Correspondingly, if assumptions in (\ref{Eq:MultivariateConditions1}) and (\ref{Eq:MultivariateConditions2}) hold true, then the second order price-dividend ratio, $\psi_{2}^{(\alpha)}(\bm{g}(k))$, can be computed as a linear system of $m^{\gamma}$ equations in $m^{\gamma}$ unknown that admits a unique solution,
    \begin{equation}\label{Eq:MultivariateSecondOrder}
        \begin{aligned}
            & r_{\alpha}^{2}\psi_{2}^{(\alpha)}(g_{a_{1}}^{(1)},\ldots,g_{a_{\gamma}}^{(\gamma)})-\\
            & \sum_{j_{1},\hdots,j_{\gamma}=1}^{m}\psi_{2}^{(\alpha)}(g_{j_{1}}^{(1)},\hdots,g_{j_{\gamma}}^{(\gamma)})(g_{j_{\alpha}}^{(\alpha)})^{2}\big(\prod_{f=1}^{\gamma}\sum_{w=1}^{\gamma}\sum_{c=1}^{m}e_{c}^{(w)}(k)\lambda_{w,f}\mathbf{P}_{c,j_{f}}^{(w,f)}\big)\\
            & =2\sum_{j_{1},\hdots,j_{\gamma}=1}^{m}\psi_{1}^{(\alpha)}(g_{j_{1}}^{(1)},\hdots,g_{j_{\gamma}}^{(\gamma)})\cdot (g_{j_{\alpha}}^{(\alpha)})^{2}\cdot \big(\prod_{f=1}^{\gamma}\sum_{w=1}^{\gamma}\sum_{c=1}^{m}e_{c}^{(w)}(k)\lambda_{w,f}\mathbf{P}_{c,j_{f}}^{(w,f)}\big)\\
            & + \sum_{j=1}^{m}\sum_{\beta=1}^{\gamma}\sum_{h=1}^{m}e_{h}^{(\beta)}(k)\lambda_{\beta,\alpha}\mathbf{P}_{h,j}^{(\beta,\alpha)}(g_{j}^{(\alpha)})^{2}.
        \end{aligned}
    \end{equation}
    
    The solutions of the first and second order price-dividend ratio in (\ref{Eq:MultivariateFirstOrder}) and (\ref{Eq:MultivariateSecondOrder}) present a different price-dividend ratio attached to each combination of states of the growth-process of each stock.
    
    Finally, considering the possible correlation between the stocks and holding assumptions in (\ref{Eq:MultivariateConditions1}) and (\ref{Eq:MultivariateConditions2}), it is possible to compute the product price-dividend ratio, $\psi_{2}^{(\alpha;\beta)}(\bm{g}(k))$,
    \begin{equation}\label{Eq:MultivariatePriceProduct}
        \begin{aligned}
            & r_{\alpha}r_{\beta}\psi_{2}^{(\alpha,\beta)}(g_{a_{1}}^{(1)},\ldots,g_{a_{\gamma}}^{(\gamma)})=\sum_{j_{\alpha},j_{\beta}=1}^{m}g_{j_{\alpha}}^{(\alpha)}g_{j_{\beta}}^{(\beta)}\big(\prod_{f\in \{\alpha,\beta\}}\sum_{w=1}^{\gamma}\sum_{c=1}^{m}e_{c}^{(w)}(k)\lambda_{w,f}\mathbf{P}_{c,j_{f}}^{(w,f)}\big)\\
            & +\!\!\!\!\!\sum_{j_{1},\hdots,j_{\gamma}=1}^{m}\!\!\!\psi_{1}^{(\beta)}(g_{j_{1}}^{(1)},\hdots,g_{j_{\gamma}}^{(\gamma)})(g_{j_{\alpha}}^{(\alpha)})(g_{j_{\beta}}^{(\beta)})\big(\prod_{f=1}^{\gamma}\sum_{w=1}^{\gamma}\sum_{c=1}^{m}e_{c}^{(w)}(k)\lambda_{w,f}\mathbf{P}_{c,j_{f}}^{(w,f)}\big)\\
            & +\!\!\!\!\!\sum_{j_{1},\hdots,j_{\gamma}=1}^{m}\!\!\!\psi_{1}^{(\alpha)}(g_{j_{1}}^{(1)},\hdots,g_{j_{\gamma}}^{(\gamma)})(g_{j_{\alpha}}^{(\alpha)})(g_{j_{\beta}}^{(\beta)})\big(\prod_{f=1}^{\gamma}\sum_{w=1}^{\gamma}\sum_{c=1}^{m}e_{c}^{(w)}(k)\lambda_{w,f}\mathbf{P}_{c,j_{f}}^{(w,f)}\big)\\
            & +\!\!\!\!\!\sum_{j_{1},\hdots,j_{\gamma}=1}^{m}\!\!\!(g_{j}^{(\alpha)})(g_{j}^{(\beta)})\psi_{1}^{(\alpha)}(g_{j_{1}}^{(1)},\hdots,g_{j_{\gamma}}^{(\gamma)})\psi_{1}^{(\beta)}(g_{j_{1}},\hdots,g_{j_{\gamma}})\big(\prod_{f=1}^{\gamma}\sum_{w=1}^{\gamma}\sum_{c=1}^{m}e_{c}^{(w)}(k)\lambda_{w,f}\mathbf{P}_{c,j_{f}}^{(w,f)}\big).
        \end{aligned}
    \end{equation}
    
    Knowing the product price-dividend ratio for any couple $(\alpha, \beta)$ of stocks, it is simple to compute the covariance function between the prices of two stocks:
    \begin{equation}\label{Eq:Covariance}
    \begin{aligned}
        &Cov(\mathcal{P}^{(\alpha)}(\bm{g}(k)),\mathcal{P}^{(\beta)}(\bm{g}(k)))\\
        &=\mathbb{E}_{(k)}[\mathcal{P}^{(\alpha)}(\bm{g}(k))\cdot \mathcal{P}^{(\beta)}(\bm{g}(k))]-\mathbb{E}_{(k)}[\mathcal{P}^{(\alpha)}(\bm{g}(k))]\cdot \mathbb{E}_{(k)}[\mathcal{P}^{(\beta)}(\bm{g}(k))]\\
        & =d^{(\alpha)}(k)d^{(\beta)}(k)\bigg(\psi_{2}^{(\alpha,\beta)}(\bm{g}(k))-\psi_{1}^{(\alpha)}(\bm{g}(k))\psi_{1}^{(\beta)}(\bm{g}(k))\bigg).
    \end{aligned}
    \end{equation}
    
    The authors apply the model to a portfolio of three US stock with a stable dividend policy with a long history and compare results with other valuation models. Finally, they show how to obtain the risk of the portfolio for different combinations of the stocks.
    
\section{Conclusion}\label{Sec:Conclusion}
    This Chapter presented a review of the dividend discount model from its basic formulation to more recent and advanced stochastic models based on the Markov chain modelling of the dividend process. As the fundamental valuation of the firms represents an important function in the financial markets, especially for long-term investments, the Markov stock model clearly show some advantages over the Gordon model and its extensions. In particular, the Markov stock model permits to obtain a different valuation depending on the state of the growth-dividend process, or on a combination of the states of the various series in the multivariate case.
    
    However, the Markov stock model presents some limitations, that are shared with the other cited models. First, the valuation is based on the dividend process, therefore it is only applicable to companies that pay dividend and with a long history of payments. Second, the discounting factor, $k_e$, is considered constant, thus it is not a realistic assumptions when considering the very long timeframe.
    
    Future extensions of the Markov stock model could consider the inclusion of some restrictions on the estimation of the transition probability matrix to reduce the number of parameters to estimate and permit the use of shorter dividend series. Moreover, the cost of equity could be modelled as a stochastic process interdependent with the dividend process. Finally, the model could be extended to companies without a dividend policy, perhaps using the earnings or similar cash flows.
    
\bibliographystyle{agsm}
\bibliography{bib}

\vfill\pagebreak
\
\thispagestyle{empty}
\end{document}